\documentclass[twocolumn, showpacs, showkeys, amsmath, amssymb, superscriptaddress]{revtex4-1}

\newif\iffinal
\finaltrue

\usepackage[normalem]{ulem} 	
\usepackage[usenames,dvipsnames]{color}
\usepackage{graphicx}   
\usepackage{xcolor} 
\usepackage{hyperref}

\newcommand{\del}[1]{\sloppy{\textcolor{blue}{\sout{#1}}}} 
\newcommand{\macom}[1]{{\marginpar{\textcolor{blue}{#1}}}} 

\iffinal

	\renewcommand{\del}[1]{}

	\renewcommand{\macom}[1]{}
	
\else
\fi

\renewcommand{\thefigure}{\textbf{\arabic{figure}}}

\begin{document}

\title{Resolving the positions of defects in superconducting quantum bits}

\author{Alexander Bilmes\**}
\affiliation{Physikalisches Institut, Karlsruhe Institute of Technology, 76131 Karlsruhe, Germany}
\author{Anthony Megrant}
\affiliation{Google, Santa Barbara, California 93117, USA}
\author{Paul Klimov}
\affiliation{Google, Santa Barbara, California 93117, USA}
\author{Georg Weiss}
\affiliation{Physikalisches Institut, Karlsruhe Institute of Technology, 76131 Karlsruhe, Germany}
\author{John M. Martinis}
\affiliation{Google, Santa Barbara, California 93117, USA}
\author{Alexey V. Ustinov}
\affiliation{Physikalisches Institut, Karlsruhe Institute of Technology, 76131 Karlsruhe, Germany}
\author{J\"urgen Lisenfeld}
\affiliation{Physikalisches Institut, Karlsruhe Institute of Technology, 76131 Karlsruhe, Germany}
	
\date{\today}
\begin{abstract}
Solid-state quantum coherent devices are quickly progressing. Superconducting circuits, for instance, have already been used to demonstrate prototype quantum processors comprising a few tens of quantum bits. This development also revealed that a major part of decoherence and energy loss in such devices originates from a bath of parasitic material defects. However, neither the microscopic structure of defects nor the mechanisms by which they emerge during sample fabrication are understood. Here, we present a technique to obtain information on locations of defects relative to the thin film edge of the qubit circuit. Resonance frequencies of defects are tuned by exposing the qubit sample to electric fields generated by electrodes surrounding the chip. By determining the defect's coupling strength to each electrode and comparing it to a simulation of the field distribution, we obtain the probability at which location and at which interface the defect resides. This method is applicable to already existing samples of various qubit types, without further on-chip design changes. It provides a valuable tool for improving the material quality and nano-fabrication procedures towards more coherent quantum circuits.
\end{abstract}

\flushbottom
\maketitle
%
%
\thispagestyle{empty}

\section*{Introduction}

Material defects have a manifold of microscopic origins such as impurities in solids or adsorbates hosted on surfaces~\cite{Muller:2017}. Their detrimental role was identified already in first experiments with superconducting quantum bits (qubits)~\cite{Simmonds04}. Strong interaction with a long-known defect type, charged two-level systems (TLS)~\cite{Anderson:PhilMag:1972,Phillips:JLTP:1972}, residing in the tunnel barrier of a qubit's Josephson junction gives rise to avoided level crossings and resonant energy absorption~\cite{Cooper04}. This form of dielectric loss could be mitigated by reducing the amount of lossy dielectrics, e.g. by incorporating smaller Josephson junctions~\cite{Martinis:PRL:2005} and by avoiding insulating layers. Another strategy is to enlarge the footprint of device capacitors in order to dilute the electric field induced by the qubit, which excites defects by coupling to their electric dipole moments~\cite{Wang:APL:2015,Gambetta17}. As a consequence of significantly enhanced coherence times, qubits became sensitive also to weakly coupling defects residing on the surfaces and interfaces of circuit electrodes~\cite{Barends13}. Since these are limiting the performance of state-of-the-art circuits~\cite{klimov2018,Schloer2019,Burnett2019}, further progress towards scaled-up quantum processors requires strong efforts to prevent the appearance of defects, e.g. by using better materials, improved fabrication procedures~\cite{Schoelkopf2016,Pappas2017,Dunsworth2017,RigettiFab}, and surface treatment to avoid contamination and parasitic adsorbates~\cite{Kumar16,deGraaf17}. This endeavor needs to be guided by careful analysis of defect properties such as densities, electric dipole moments, and positions, in order to identify and improve the manufacturing steps that reduce defect formation, and to analyze the microscopic structure of defects.\\

In this Letter, we present a method to extract information on the spatial positions of defects at the profile of the film edge in a qubit circuit. For doing this, we exploit the tunability of a charged defect's resonance frequency $\omega$ by an electric field $\mathbf{E}$,
\begin{equation}
\omega = \sqrt{\mathnormal{\Delta}^2 + \varepsilon^2}/\hbar, \ \ \ \ \varepsilon = \varepsilon_i + 2\mathbf{p}\mathbf{E},
\label{eq:TLSfreq}
\end{equation}
where $\hbar$ is the reduced Planck constant, and $\mathbf{p}$ is the defect's electric dipole moment. The offset energy $\varepsilon_i$ is given by local strain and electric fields from surrounding atoms, and $\mathnormal{\Delta}$ is the defect's (constant) tunneling energy~\cite{Anderson:PhilMag:1972,Phillips87}. In our experiment, $\mathbf{E}$ is composed of electric fields $\mathbf{E}_\text{t}$ and $\mathbf{E}_\text{b}$ generated by two gate electrodes placed above (t) and below (b) the sample chip, respectively. The qubit is used to monitor the defect's resonance frequency and their responses to DC voltages $V_\text{t}$ and $V_\text{b}$ applied to the respective top and bottom electrode.
Comparing the response to the spatial variation of the applied electric fields obtained from finite element simulations, the position $\mathbf{x}$ of a defect can be deduced by solving the equation
\begin{equation}
	\gamma_\text{t} V_\text{t}/\gamma_\text{b} V_\text{b} = \mathbf{p}\mathbf{E}_\text{t}(V_\text{t},\mathbf{x})/\mathbf{p}\mathbf{E}_\text{b}(V_\text{b},\mathbf{x}).
	\label{eq:ratio}
\end{equation}
Here, $\gamma_\text{t}$ and $\gamma_\text{b}$ are the defect's tunability coefficients by the respective top and bottom fields, which are obtained by fitting the measured resonance frequency dependence of each defect
\begin{equation}
\omega=\sqrt{\mathnormal{\Delta}^2+\left( \varepsilon_\text{i} + \gamma_\text{t} V_\text{t} + \gamma_\text{b} V_\text{b} \right) ^2}/\hbar
\label{eq:hyperbola}
\end{equation}\\
derived from Eq.~\eqref{eq:TLSfreq} using the identities $2\mathbf{pE}_\text{t}=\gamma_\text{t} V_\text{t}$ and $2\mathbf{pE}_\text{b}=\gamma_\text{b} V_\text{b}$. Since DC electric fields approach metallic electrodes always perpendicular to their surface, Eq.~(\ref{eq:ratio}) can be reduced by dropping the dipole moment projections onto each field, and regarding the absolute field values. Thus, at metal film interfaces the defect locations are deduced from the simplified equation $\gamma_\text{t}/\gamma_\text{b}=|\mathbf{E}_\text{t}|/|\mathbf{E}_\text{b}|$.
At the substrate-vacuum interface however, the applied fields are not necessarily parallel, which requires accounting for the defect's dipole moment orientation, as described below.\\
\begin{figure}[htbp]
	\centering
	\includegraphics[width=\columnwidth]{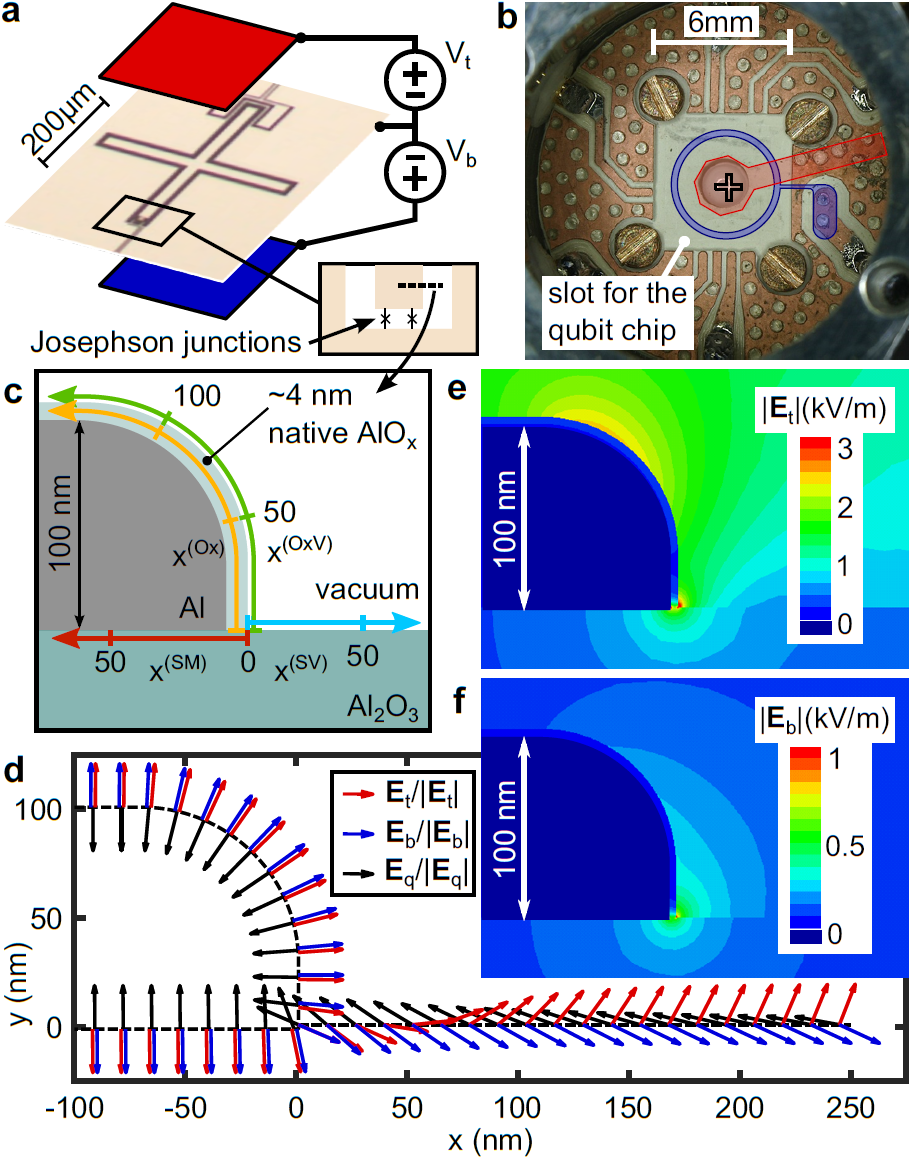}
	\caption{\textbf{Experimental setup for E-field tuning of defects and simulation results.} \textbf{a} Qubit picture with sketched top (red) and bottom (blue) electrodes which are controlled by two independent voltage sources referenced to the on-chip ground plane. \textbf{b} Photograph of the opened sample housing without qubit chip, and illustration of the top (red) and annular bottom electrode (blue). The cross mark denotes the location of the investigated qubit. \textbf{c} Sketch of a cross-section through the substrate and the qubit electrode near the film edge, showing the interfaces of interest: the substrate-metal (SM) and substrate-vacuum (SV) interfaces, the native AlO$_x$ layer (Ox), and the oxide-vacuum interface (OxV). Colored arrows define the coordinate systems along each interface, which have their origins close to the substrate-metal-vacuum edge. \textbf{d} Directions of the simulated DC fields $\mathbf{E}_\text{t}$ and $\mathbf{E}_\text{b}$ generated by top and bottom electrodes, and of the qubit's plasma oscillation field $\mathbf{E}_\text{q}$. \textbf{e},\textbf{f} Simulated electric field strengths for a voltage of $-0.5\,\mathrm{V}$ applied to either top or bottom electrodes.}
	\label{fig:1}
\end{figure}
\section*{Experiment}
We connect each of the two electrodes to a DC voltage source that is referenced to the common ground of the sample housing and the on-chip groundplane as sketched in Fig.~\ref{fig:1}\textbf{a}.  A photograph of the opened aluminum sample housing is shown in Fig.~\ref{fig:1}\textbf{b}, where red and blue structures indicate the real dimensions of the DC electrodes. The top electrode consists of a copper/Kapton-foil sandwich glued to the lid of the sample housing. The bottom electrode is integrated in the PCB backplane. Its circular shape allows a piezo-mechanical transducer to exert mechanical force onto the center of the qubit chip, which allows one to tune defects by elastic strain~\cite{Lisenfeld2015, Grabovskij12}. In this work, the piezo is not used, and we refer to Ref.~\cite{Lisenfeld19} for a study comparing the defect response to mechanical strain and electric fields that was performed with the same setup. Further technical details of the sample housing are provided in Supp. Mat. 1.\\

The employed qubit sample is an aluminum-based transmon qubit~\cite{KochTransmon} consisting of a cross-shaped capacitor electrode that is connected to ground by a split Josephson junction~\cite{Barends13}. The applied E-field is expected to be constant along the film edges of the qubit island and of the surrounding groundplane due to the qubit's geometric symmetry and its central position relative to the electrodes (see Supp. Mat.~1). Accordingly, the problem to simulate the spatial dependence of the electric field can be limited to a 2-dimensional cross-section focusing on the film edge where the fields are strongest. This region is illustrated in Figure~\ref{fig:1}\textbf{c}, labeling the interfaces of interest which are the substrate-vacuum (SV), and the three film interfaces: substrate-metal (SM) interface, the inside of the native aluminum oxide layer (Ox), and the oxide-vacuum interface (OxV). The colored arrows indicate the spatial coordinates along each interface that have their origins at the substrate-metal-vacuum edge. As confirmed by SEM investigation, the edge of the film is rounded because it was patterned using isotropic reactive ion etching.\\

The E-field distribution that results from simulations is shown in Figs.~\ref{fig:1}\textbf{d}-\textbf{f}. In Fig.~\ref{fig:1}\textbf{d}, the direction of applied fields and of the qubit's AC electric field $\mathbf{E}_\text{q}$ at the different circuit interfaces are indicated by colored arrows. Figures~\ref{fig:1}\textbf{e} and \textbf{f} show the electric field strengths generated by applying a voltage of $-0.5\,\mathrm{V}$ to one of the electrodes. These show that the top and bottom fields are supposed to be focused in different regions, which is the key point to resolve locations of defects from their response to each gate electrode. Furthermore, the qubit AC field is concentrated in the same region as the DC fields, which implies that all defects residing at the investigated interfaces and detectable by the qubit couple to the global electrodes, as detailed in Supp. Mat.~1.\\
\begin{figure}[htbp]
	\centering
	\includegraphics[width=\columnwidth]{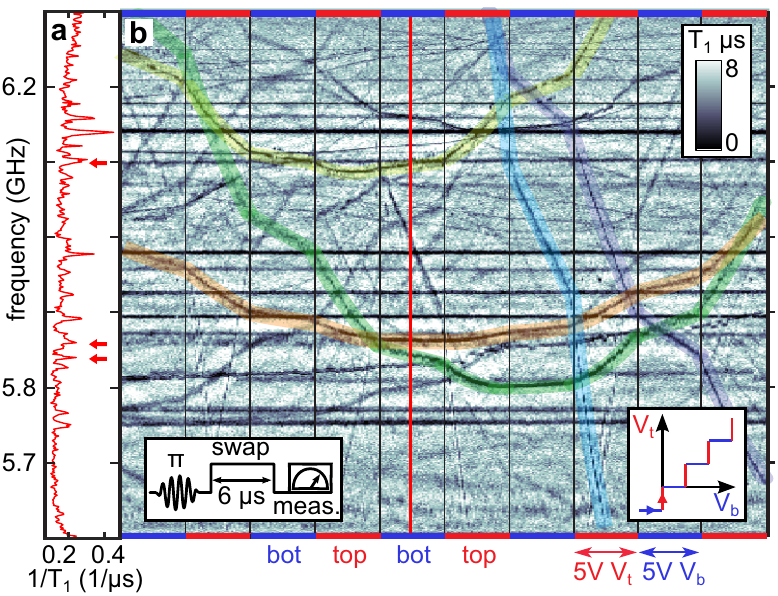}
	\caption{\textbf{Tuning defects by electric fields.}
	\textbf{a} Frequency-dependent energy relaxation rate $1/T_1$ of the qubit, measured by a swap spectroscopy protocol sketched in the left inset of \textbf{b}. Distinct peaks stem from resonant interaction with individual defects. \textbf{b} Dependence of defect resonance frequencies (dark traces) on applied voltage ramps alternating the top (red margins) and the bottom electrode (blue margins). The right inset contains the underlying ramp path in the voltage plane. Colored curves in the main figure highlight hyperbolic fits to Eq.~(\ref{eq:hyperbola}) with asymptotic slopes $\gamma_\text{t/b}$ which reflect the asymmetry tunability of defects by each electrode, characteristic of their position. The red vertical line indicates the exemplary trace shown in \textbf{a} where red arrows denote resonances of the highlighted defect traces. Horizontal black lines stem from defects hosted in the qubit's stray Josephson junction as explained in a previous work~\cite{Lisenfeld19}.} 
	\label{fig:2}
\end{figure}

The resonance frequencies of defects are detected by qubit swap spectroscopy~\cite{Cooper04,Lisenfeld2015} using the protocol shown in the left inset of Fig.~\ref{fig:2}\textbf{b}. This provides a measurement of the qubit's energy relaxation rate $1/T_1$ whose frequency dependence displays Lorentzian peaks due to resonant interaction with defects, as visible in Fig.~\ref{fig:2}\textbf{a}. To obtain the defects' response coefficients $\gamma_\text{t}$ and $\gamma_\text{b}$ required to solve Eq.~(\ref{eq:ratio}), we alternate measurements where the voltage on either the top or the bottom electrodes is swept upwards. The right inset of Fig.~\ref{fig:2}\textbf{b} illustrates such a sweeping path in the $V_\textbf{t}$-$V_\textbf{b}$ space, and the main figure shows resulting data. Segmented hyperbolic traces of individual defects exhibit unequal slopes in reaction to the two gates. According to Figs.~\ref{fig:1} \textbf{e} and \textbf{f}, these slopes reflect different local field strengths $\mathbf{E}_\text{t}$ and $\mathbf{E}_\text{b}$. A few exemplary fits to Eq.~\eqref{eq:hyperbola} are indicated by highlighted curves in Fig.~\ref{fig:2}\textbf{b}. Horizontal traces indicate defects which most probably reside inside the qubit's stray Josephson junction
where they do not experience any applied DC fields, as further detailed in the previously mentioned work~\cite{Lisenfeld19}.\\

\begin{figure*}[htbp]
	\includegraphics[width=\textwidth]{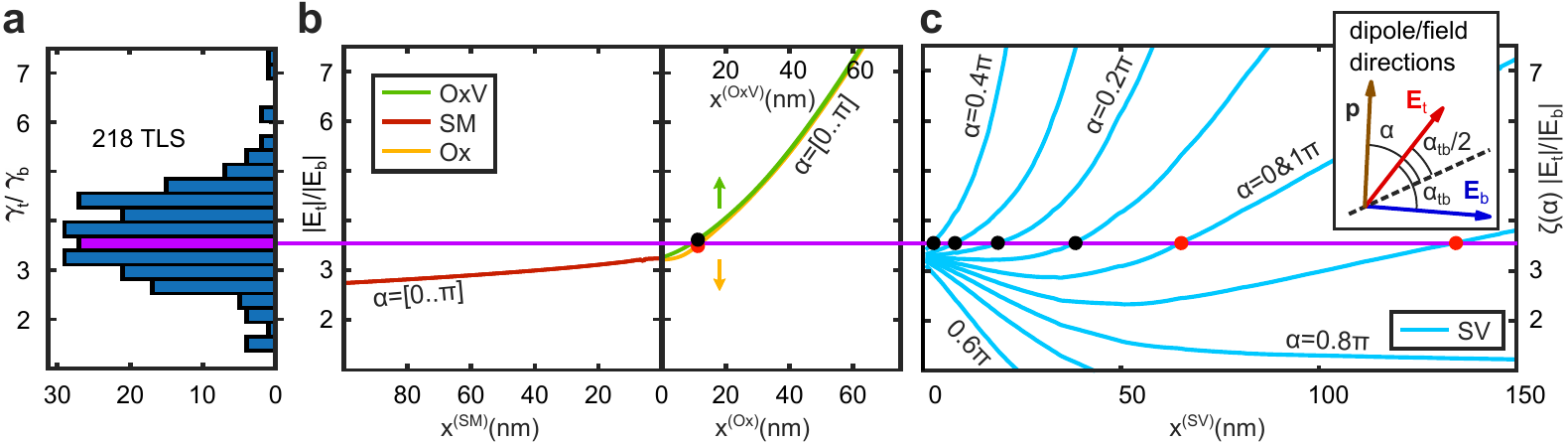}
	\caption{\textbf{Deduction of defect positions.}
	 \textbf{a} Measured ratios $\gamma_\text{t}/\gamma_\text{b}$ of defect tunability coefficients by the top and bottom electrodes. \textbf{b} Ratio of the simulated electric field strengths $|\mathbf{E}_\text{t}|/|\mathbf{E}_\text{b}|$ along the different interfaces as indicated by the axes labels. Since the applied DC fields are parallel at these interfaces, the field ratio does not depend on $\alpha$ which defines the defect's dipole moment orientation (see inset of \textbf{c}). \textbf{c} Right side of Eq.~\eqref{eq:ratio} for some $\alpha$ values, where $\zeta(\alpha,\mathbf{x})\equiv\cos(\alpha-\alpha_\textbf{tb}(\mathbf{x})/2)/\cos(\alpha+\alpha_\textbf{tb}(\mathbf{x})/2)$ is the ratio of dipole moment projections onto the fields, and $\alpha_\textbf{tb}$ is the angle between $\mathbf{E}_\text{t}$ and $\mathbf{E}_\text{b}$. The horizontal violet line through \textbf{b} and \textbf{c} provides graphical solutions of Eq.~\eqref{eq:ratio} for possible positions of an exemplary defect (black dots), of which those requiring nonphysically large electric dipole moments $>10\,\mathrm{Debye} \approx 2 e \mathrm{\AA}$ are discarded (red points).}
	\label{fig:3}
\end{figure*}

A distribution of measured $\gamma_\text{t}/\gamma_\text{b}$ ratios is plotted in Fig.~\ref{fig:3}\textbf{a}, comprising data from 218 field-dependent defects that could be detected with this qubit in a frequency range of $0.7\,\mathrm{GHz}$ (between $5.6\,\mathrm{GHz}$ and $6.3\,\mathrm{GHz}$) and an applied voltage range of $V_\text{t},V_\text{b} \in [-100..100]\,\mathrm{V}$. As expected, all defects are more strongly tuned by the top electrode that induces a larger field at the qubit position, due to its geometry (cf. Fig.~\ref{fig:1}\textbf{b}).\\
\section*{Results}

As stated before, defect locations at film interfaces (SM, Ox, and OxV) are given by solutions of $\gamma_\text{t}/\gamma_\text{b}=|\mathbf{E}_\text{t}|/|\mathbf{E}_\text{b}|$. The ratio of simulated electric field strengths is shown in Fig.~\ref{fig:3}\textbf{b} along the respective coordinates defined in Fig.~\ref{fig:1}\textbf{c}. The horizontal violet line through Fig.~\ref{fig:3}\textbf{b} provides graphical solutions of Eq.~\eqref{eq:ratio} for an exemplary defect with measured ratio $\gamma_\text{t}/\gamma_\text{b}\approx3.5$. The two solutions at the Ox and OxV interfaces are indicated by red and black dots, respectively, and correspond to a defect's distance of $\sim 15\,\mathrm{nm}$ from the
substrate-metal-vacuum edge. Since the applied fields are parallel at film interfaces, the solutions are degenerate in the dipole moment orientation $\alpha\in[0..\pi]$ defined in the inset of Fig.~\ref{fig:3}\textbf{c}. At the SV interface, the tunability ratio depends in addition on $\alpha$, as visualized in Fig.~\ref{fig:3}\textbf{c} where the right term of Eq.~\eqref{eq:ratio} is plotted for some $\alpha$ values. Hereby, $\zeta(\alpha,\mathbf{x})\equiv\cos(\alpha-\alpha_\text{tb}(\mathbf{x})/2)/\cos(\alpha+\alpha_\text{tb}(\mathbf{x})/2)$ is the ratio of dipole moment projections on the fields, and $\alpha_\text{tb}$ is the angle between $\mathbf{E}_\text{t}$ and $\mathbf{E}_\text{b}$. The red and black dots indicate anticipated locations of the exemplary defect at the SV interface.\\



The defect's electric dipole moment component $p_\parallel$ parallel to the applied fields is calculated from each solution $\mathbf{x}$ and a corresponding field strength $\mathbf{E_\textbf{t/b}}(\mathbf{x})$, as reported in Supp. Mat. 1. Inside the oxide layer (Ox), electric fields are reduced by about a factor of 10 due to the materials' permittivity. A possibility for the exemplary defect ($\gamma_\text{t}=102\,\mathrm{\hbar MHz/V}$, $\gamma_\text{b}=29\,\mathrm{\hbar MHz/V}$) to reside at this interface would require an electric dipole moment of $p_\parallel \approx 30\,\mathrm{Debye}$, which is unrealistic. Therefore, we discard in our analysis all solutions that imply dipole moments $p_\parallel > 10\,\mathrm{Debye} \approx 2 e \mathrm{\AA}$ which is a maximum imaginable number in solids~\cite{Yu2020}. The cutoff value is discussed in Supp. Mat.~3. In Figs.~\ref{fig:3}\textbf{b} and \textbf{c}, valid and truncated solutions from the current example are indicated by black and red dots, respectively.
Regarding possible locations at the SV interface, only angles $\alpha \in[0.1..0.4]\,\mathrm\pi$ lead to reasonable dipole moment sizes for this defect, corresponding to distances $x^\text{SV} \in 
[40..5]\,\mathrm{nm}$ from the substrate-metal-vacuum edge. The probability for a given defect to reside at a particular interface is defined by the range of allowed $\alpha$. Here, $\alpha$ spans over $0.4\,\mathrm{\pi}$ at the SV, and over $\mathrm{\pi}$ at the OxV interfaces, resulting in $P_\mathrm{SV} = 0.4 \pi / (\pi + 0.4 \pi)\approx 0.29$ and $P_\mathrm{OxV} = \pi /  (\pi + 0.4 \pi)\approx 0.71$, respectively. This calculation is repeated for each field-tunable defect using hundreds of interpolated $\alpha$ values to increase the probability precision.\\
\begin{figure}[htbp]
	\centering
	\includegraphics[width=\columnwidth]{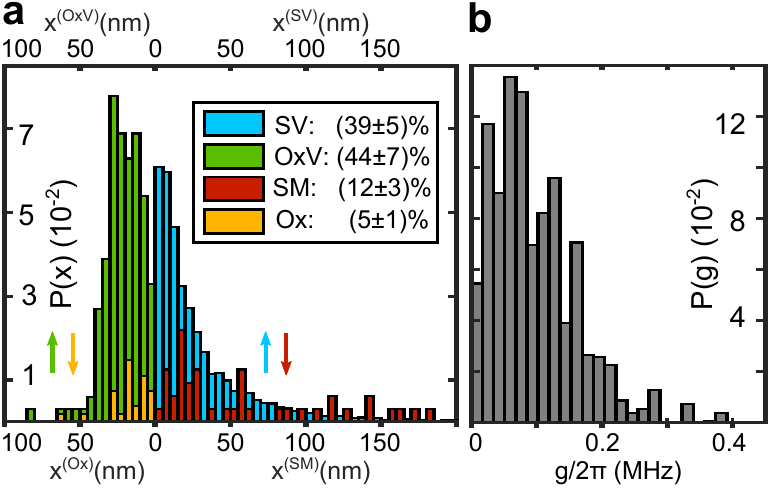}
	\caption{\textbf{Defect positions and coupling strengths.}
	\textbf{a} Histograms of deduced defect locations, whose relative weights reveal the probability of finding a field-tunable defect at the respective interface (see legend). The probability errors are deduced from the estimation error of the electrode distances to the qubit film as detailed in Sup.~3. The interface to air (OxV, SV) hosts most detectable defects, possibly due to fabrication contaminants and adsorbates. Further, detected defects are concentrated within $50\,\mathrm{nm}$ from the substrate-metal-vacuum edge since the qubit fields are focused in this region. \textbf{b} Distribution of extracted defect-qubit coupling strengths $g$, which is in good agreement to direct measurements using the same sample~\cite{Lisenfeld19}.
	}
	\label{fig:4}
\end{figure}
The main result of this work is presented in Fig.~\ref{fig:4}\textbf{a}, showing the histogram of extracted positions of field-tunable defects weighted by the average probability to reside at a certain interface (colors). Most of such defects are located within a distance of about $50\,\mathrm{nm}$ to the substrate-metal-vacuum edge ($x=0$). This is expected since the qubit AC-electric field is mainly concentrated in this region. More distant defects couple too weakly to be detected with the given qubit coherence time. With highest probability, defects reside at open surfaces of the sample, i.e. at the OxV and SV interfaces. Due to the reduced strength of applied fields inside the native oxide, most solutions at the Ox interface resulted in unrealistic dipole moment sizes, and were discarded. The SV histogram appears particularly broad and smoothed due to tracing out the dipole moment orientation $\alpha$ which is assumed isotropic.\\

Once the defect positions are identified, we can calculate their coupling strengths to the qubit $g = \mathbf{pE}_\text{q}(\mathbf{x})$ from the deduced dipole moment projections on the applied fields and the simulated qubit AC field $\mathbf{E}_\text{q}$. Figure~\ref{fig:4}\textbf{b} shows the resulting distribution of $g$, which is very similar to an independent and direct measurement of the coupling strengths reported in Ref.~\cite{Lisenfeld19}.\\

We note that our analysis does not resolve defect locations along the qubit film edge, especially it disregards the possibility that defects reside on the narrow electrodes of the qubit's split Josephson junction. However, this region may be particularly critical because of the concentrated qubit field, and the additional lithography steps required to deposit sub-micron junctions, which may promote defect formation by substrate surface amorphisation and processing residuals~\cite{Dunsworth2017}. Nevertheless, we expect that the field strength ratio remains unchanged at the edges of junction-forming films so that the analysis remains valid in this region.
\\

Overall, in this sample, we detected on average 16 defects per GHz hosted in the Josephson junctions, and 26 field-tunable defects per GHz at any applied electric field. Further 5 field-tunable defects per GHz could not be located since some defects appeared only in one data segment (red and blue framed windows in Fig.~\ref{fig:2}\textbf{b}). This effect can be minimized by choosing sufficiently narrow segments. Assuming an equal distribution of field-tunable defects along the 3mm-long edge of the qubit film, we obtain a density of $\approx 10$ defects per (GHz $\cdot$ mm). We note that the sample was stored for several years while it was covered by photoresist. Incorporation of resist atoms and residuals due to inadequate cleaning may explain the degradation of qubit coherence caused by an increased number of surface defects detected in this work.\\

\section*{Conclusions}

The demonstrated technique to determine the position of defects on the surface of a quantum circuit provides a viable tool to verify the material quality and to optimize micro-fabrication steps. Our technique requires only few externally placed gate electrodes and is thus applicable directly to existing qubit chips. Since the junction's tunnel barrier is free of applied DC electric fields, junction defects can be easily identified by zero field-tunability. While this technique presumably applies to all qubit geometries built in a coplanar architecture, the analysis of our sample is certainly least time-consuming due to its geometric symmetry. This method can be further improved by performing independent measurements of the qubit-defect coupling strengths $g$ and thus their effective dipole moment sizes~\cite{Lisenfeld19}, hereby reducing uncertainties concerning the interface at which a defect resides.\\

We have found that $46\,\%$ of all defects reside on the surface of the qubit sample, and $10\,\%$ are hosted inside the native oxide or at the metal-substrate interface. The location of another $10\,\%$ could not be resolved, and $34\,\%$ of defects reside in the tunnel barrier of the qubit's stray Josephson junctions. Hereby, the amount of defects in the weak junctions is negligible. While the redundant stray junctions can be simply shorted \cite{Dunsworth2017}, qubit surfaces remain another dominant and inherent source of dielectric losses. Decreasing this loss requires rigorous studies of fabrication processes and surface treatments during or after sample fabrication in order to improve the surface quality. The reported technique can be used to examine dielectric losses at film edges in suspended or trenched qubit samples, where electric fields can be more diluted while, on the other side, a larger surface for hosting adsorbates is available. In future experiments, additional on-chip gate electrodes can provide enhanced spatial resolution helping to distinguish defects in the immediate vicinity of the tunnel junctions.\\

\subsection*{Methods}

The supplementary material contains details on the experimental setup, simulations of the electric fields, plots of complete measured data sets, and the analysis method with error estimation. Further details can be found in the PhD thesis of AB~\cite{Bilmes19}.

\subsection*{Data availability}

The supplementary material contains details on the experimental setup, simulations of the electric fields, plots of complete measured data sets, and the analysis method with error estimation. The datasets generated and analysed during the current study are available from AB on request.

\bibliography{TLS_Localization_Biblio}

\subsection*{Acknowledgements}

This work was supported by Google inc. AB acknowledges support from the Helmholtz International Research School for Teratronics (HIRST) and the Landesgraduiertenförderung-Karlsruhe (LGF). JL gratefully acknowledges funding from the Deutsche Forschungsgemeinschaft (DFG), grant LI2446-1/2. AVU acknowledges partial support by the Ministry of Education and Science of the Russian Federation in the framework of the Program to Increase Competitiveness of the NUST. MISIS (contract No. K2-2017-081) is also acknowledged. We acknowledge the support by the KIT-Publication Fund of the Karlsruhe Institute of Technology.

\subsection*{Author contributions}

The experiments were devised by AB and JL. AB implemented the experimental setup, performed E-field simulations, and analyzed the data. JL acquired the data and wrote the manuscript with AB. The qubit sample was developed and fabricated by partners from Google. All authors discussed the results and commented on the manuscript.

\subsection*{Competing interests}

The authors declare no competing interests.

\clearpage

\onecolumngrid
\appendix
\renewcommand{\thefigure}{\textbf{S\arabic{figure}}}
\setcounter{figure}{0}

\begin{center}
\large
\textbf{Resolving the positions of defects in superconducting quantum bits\\}
\vspace{0.3cm}
\textbf{Supplementary Material}
\normalsize
\end{center}

\subsection{Experimental setup, qubit sample, and E-field simulations}
\label{SM:simu}
The sample measured in this work was fabricated by Barends et al. as described in Ref.~\cite{Barends13}. This chip contained three uncoupled transmon qubits in 'Xmon' geometry, each consisting of a cross-shaped capacitor electrode connected to ground via two Josephson junctions in parallel. Between sample fabrication and our measurements, about four years have past during which the sample was covered with photoresist which might not have been completely removed prior to our measurements. We thus note that incorporation of contaminants from the resist and its residuals may enhance the number of surface defects detected in our experiments.\\

\begin{figure*}[htb]
	\includegraphics{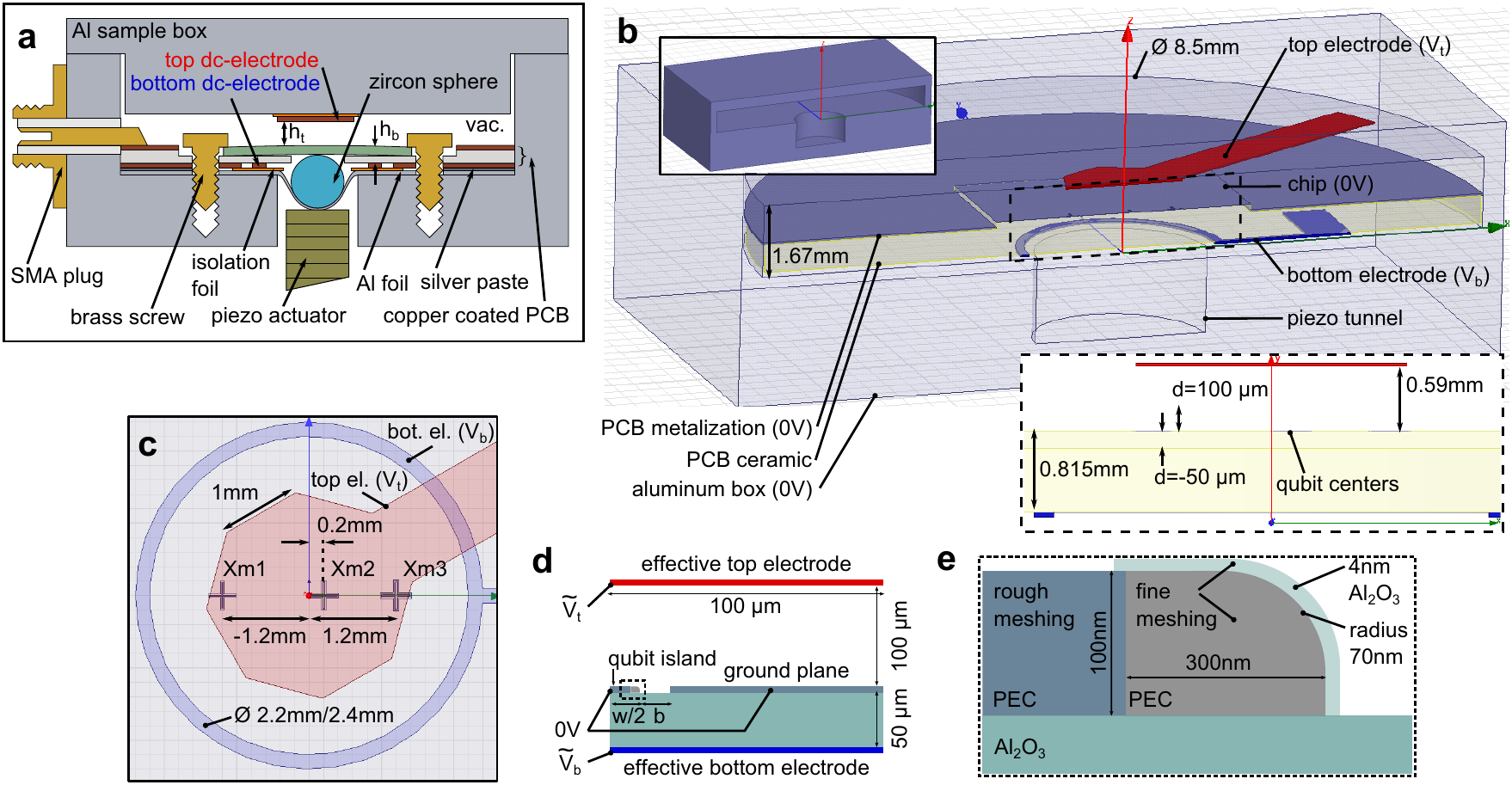}
	\caption{\textbf{Setup and simulations for electric field tuning of defects.} \textbf{a} Sketched cross-section of the sample housing. The piezo actuator is not necessary for this experiment. However, it was useful to obtain independent data sets in the same cool down, as explained in Fig.~\ref{fig:gamma_dpar_vs_gamrat}. \textbf{b} Model for coarse field simulations created with ANSYS Maxwell. The main plot shows a cut-away of the sample cavity which contains a printed circuit board (PCB, light yellow) carrying planar microwave lines* to which the sample chip is wire-bonded. The sample housing contains a circular hole in its center (piezo tunnel) allowing a piezo actuator* to exert force onto the qubit chip in order to tune defects by mechanical strain as employed in Ref.~\cite{Lisenfeld19}. Therefore, the bottom electrode below the chip has a circular shape. The top electrode consists of a copper foil/Kapton foil sandwich that is glued to the lid of the sample housing. (*) not contained in the simulation model. \textbf{c} shows the exact dimensions of the DC-electrodes from a top view. The bottom electrode is ring-shaped to leave space for the zircon sphere which mediates the elastic strain applied to the chip using the piezo actuator. The grey crosses indicate the positions of the 3 Xmon qubits on the chip. In this work, we only discuss data obtained on the centre qubit (Xm2) for which the electric fields are constant along the edge of the qubit islands. \textbf{d} Reduced 2D-model to enhance the precision of field simulations. The electrodes are modelled by effective parallel plates located 50 and 100 $\mu$m below and above the chip, respectively. These are biased at reduced voltages $\widetilde V_\text{t/b}$ so that the electric fields have equal strengths as those generated by the real electrodes, which we find from the full 3D simulation. The qubit island potential is set to ground due to the transmon regime of the qubit. \textbf{e} Magnified view of the film edge profile. Only a small margin of the edge cross-section
	is fine resolved (maximum mesh width $0.1\,\mathrm{nm}$) since the qubit fields are concentrated at	the film edge, and defects residing further away are not detectable by the qubit. The rest of the model is automatically meshed with a maximum mesh width of 250 nm. The aluminum film is modelled as a perfect conductor (PEC), and the amorphous native oxide on top of the aluminum is represented by a 4 nm thick sapphire film.}
	\label{fig:simus}
\end{figure*}

\begin{figure*}[htb]
  	\includegraphics[scale=1.5]{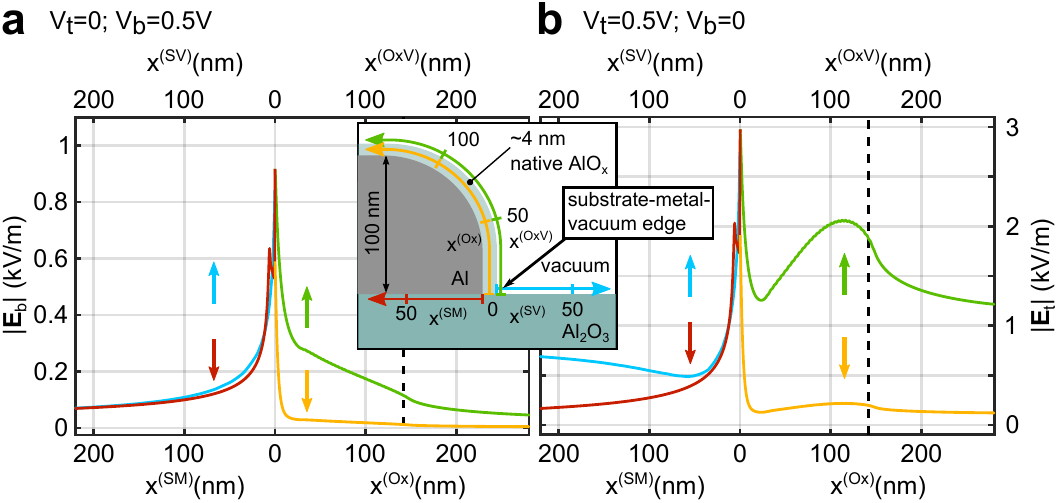}
  	\caption{\textbf{Results from E-field simulations.} Strength of the electric field generated by applying a voltage of 0.5 V to either \textbf{a} the bottom electrode or \textbf{b} the top electrode, along the substrate-vacuum (SV) interface (blue), the oxide surface (OxV, green), the substrate-metal interface (SM, red), and the inside of the oxide layer (Ox, yellow). The spacial axes along the interfaces are indicated by arrows of corresponding colors. The inset illustrates the coordinates along the different interfaces, which have their origins at the substrate-metal-vacuum edge.}
  	\label{fig:moresimus}
  \end{figure*}

We use a standard scheme for qubit readout based on detection of the state-dependent dispersive frequency shift of a readout resonator coupled to each qubit as described in the supplementary material to Ref.~\cite{Lisenfeld19}. The electric field is generated by two electrodes located above and below the qubit chip, which are both connected to independent voltage sources that are referenced to the cryostat body and on-chip grounds. Figure~\ref{fig:simus}\textbf{a} shows a sketch of the sample housing's cross-section, and Figure~\ref{fig:simus}\textbf{b} shows the three-dimensional (3D) model used for simulations of the generated electric field. The dimensions of the employed DC-electrodes are shown in Fig.~\ref{fig:simus}\textbf{c}. Fields are simulated with finite element solver ANSYS Maxwell 2015 (release 16.2.0). Once the electric fields are characterized in a full 3D-simulation, we employ the spatially reduced 2D model illustrated in Fig.~\ref{fig:simus}\textbf{d} in which the real electrodes are replaced by effective ones and a much denser meshing grid is used. Fig.~\ref{fig:simus}\textbf{e} shows dimensions of the modelled film edge, and Figure~\ref{fig:moresimus} shows the magnitudes of electric fields along the different circuit interfaces obtained from this simulation.\\

\begin{figure*}
	\includegraphics[width=0.75\columnwidth]{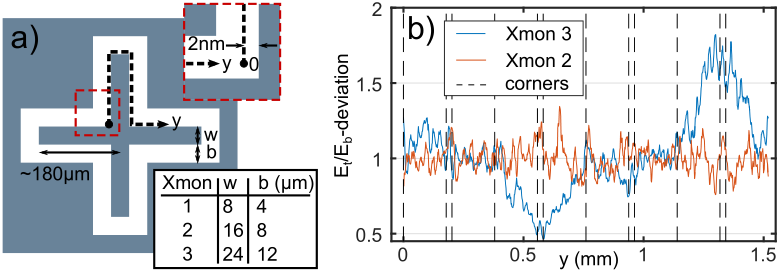}
	\caption{\textbf{Homogeneity of the electric field.} \textbf{a} Illustration of the Xmon qubit geometry. The dashed line shows the path along which the homogeneity of the electric field is verified. \textbf{b} Deviation of the electric field strength ratio $|\boldsymbol{E}_\text{t}/\boldsymbol{E}_\text{b}|$ relative to the mean value, plotted along the dashed path in \textbf{a}. The electric field was simulated in the 3D-model using a coarse resolution to save calculation time, which explains the plot roughness. For the center qubit (Xmon2, red line), the field ratio is constant along the qubit edge, in contrast to the lateral qubits. The dashed vertical lines indicate the positions of corners of the cross-shaped qubit electrode.}
	\label{fig:homogeneity}
\end{figure*}

Although we acquired data from all three qubits, in our article we only discuss results obtained on the qubit that is located in the centre of the sample chip (Xmon 2),  because only for this qubit the applied electric field is sufficiently homogeneous along the edge of the qubit island and of the surrounding ground plane. Figure~\ref{fig:homogeneity} illustrates this spatial dependence of the ratio of electric fields generated by top and bottom electrodes. In future experiments, the field homogeneity can be enhanced by using electrodes that are larger than the chip, allowing one to measure many qubit samples in the same cool-down. \\

The simulations show that the electric field induced by qubit oscillations is concentrated at film edges and decays as $\sim 1/\sqrt{x}$, where $x$ is the distance to the film edge. Hereby, a defect having an electric dipole moment of $10\,\mathrm{D}$ (maximum imaginable value in solids) and residing at $x>200\,\mathrm{nm}$ would couple to the qubit by $g/2\pi<0.05\,\mathrm{MHz}$, which is about the typical detection sensitivity of the sample. Thus, $200\,\mathrm{nm}$ is the maximum distance to the edge where defects are still detectable. Further, simulations show that within such a distance, and on both sides of the film, the dc-electric field induced by each separate global electrode exceeds $130\,\mathrm{V/m}$ per volt applied (the maximum applied voltage is $\pm100\,\mathrm{V}$). By this minimum field, the asymmetry energy of a defect having a small parallel electric dipole moment component of 1 Debye would be tuned by 130 MHz. Thus, we can be sure that every detectable defect is also clearly tunable by any of the two employed global electrodes.\\

\subsection{Defect spectroscopy}
Defects are detected by recording the frequency-dependence of the qubit energy relaxation rate which displays Lorentzian peaks due to dissipation from resonant defect interaction~\cite{Cooper04,Lisenfeld2015}. To minimize the measurement time, we employ a swap spectroscopy protocol and deduce the energy relaxation rate from a single measurement at each frequency and additional reference measurements. More details on the employed methods can be found in the supplementary material to Ref.~\cite{Lisenfeld19}.\\

\subsection{Error estimation}
\label{SM:error}
\begin{figure*}
	\includegraphics[width=\textwidth]{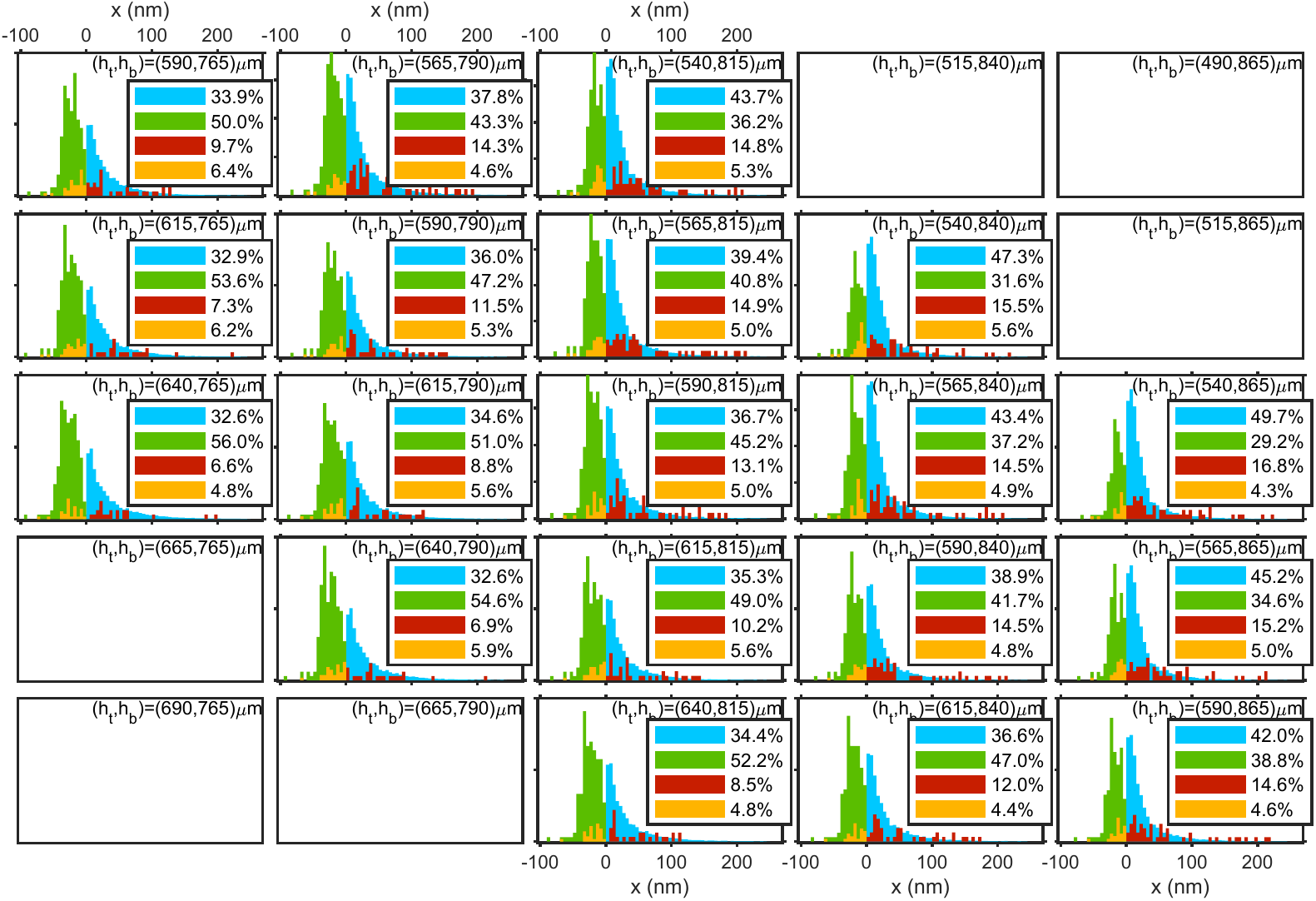}
	\caption{\textbf{Error estimation.} Defect locations at different interfaces (colors) as deduced from analysis runs for varying distances $h_\text{t}$ and $h_\text{b}$ between the qubit and the top and bottom electrodes around their nominal values $h_\text{t} = 590\,\mathrm{\mu m}$ and $h_\text{b} = 815\,\mathrm{\mu m}$, and with a variation step of $25\,\mathrm{\mu m}$. The legends indicate the participation of the given interface to the total amount of detected field-tunable defects. The interface participations and errors quoted in Fig.~4 \textbf{a} of the main text result from the arithmetic mean and standard deviation of equivalent results obtained with a smaller variation step of $10\,\mathrm{\mu m}$.}
	\label{fig:error}
\end{figure*}

\noindent\textbf{Electrode distance}\\ We estimate that the largest error in the deduced defect locations stems from uncertainties in the vertical distances $h_\text{t}$ and $h_\text{b}$ between the qubit and the top and bottom electrodes, respectively. Deviations in $h_\text{b}$ may result from uneven machining of the PCB ceramic, while $h_\text{t}$ depends on the thicknesses of insulation foils underneath the qubit chip and between the top electrode and sample housing. To estimate their influence on the results, we repeat our analysis for systematically varied electrode distances within a range of $\pm50\,\mathrm{\mu m}$ (the estimated maximum error) around their nominal values $h_\text{t} = 590\,\mathrm{\mu m}$ and $h_\text{b} = 815\,\mathrm{\mu m}$. This results in small differences for deduced defect locations and interface participations whose arithmetic mean values and standard deviations are quoted in the legend of Fig.~4 \textbf{a} of the main text. While the numbers presented in Fig.~4 \textbf{a} have been deduced from a sweep using a step of $10\,\mathrm{\mu m}$, an exemplary result using a coarse variation step of $25\,\mathrm{\mu m}$ is presented in Fig.~\ref{fig:error} to give an impression how the analysis outcome behaves under small distance variations. In each subplot, the underlying $(h_\text{t},h_\text{b})$ combination is indicated while the empty frames denote combinations outside of the allowed limit of $\pm50\,\mathrm{\mu m}$. Further, the distance variations obey the fact that $h_\text{t}$ depends on changes in $h_\text{b}$ due to the sample holder geometry (see Fig.~\ref{fig:simus}\textbf{a}), but not vice versa. For example, if we choose $h_\text{t}$ to be shortened by $50\,\mathrm{\mu m}$ and $h_\text{b}$ by $25\,\mathrm{\mu m}$, the resulting distances will be $(h_\text{t}-50\,\mathrm{\mu m}+25\,\mathrm{\mu m},h_\text{b}-25\,\mathrm{\mu m})=(565,790)\,\mathrm{\mu m}$.\\

\noindent\textbf{Film edge shape}\\ Deviations in the shape of the qubit electrode from the assumed rounded profile, e.g. due to fabrication imperfections or local film corrosion, will result in local electric field strengths that differ from the simulation results. However, since our analysis is based on the comparison of the strengths of E-fields induced by the top and bottom electrodes, which presumably are equally affected by film imperfections, we expect that the analysis remains valid.\\

\begin{figure}
	\centering
	\includegraphics{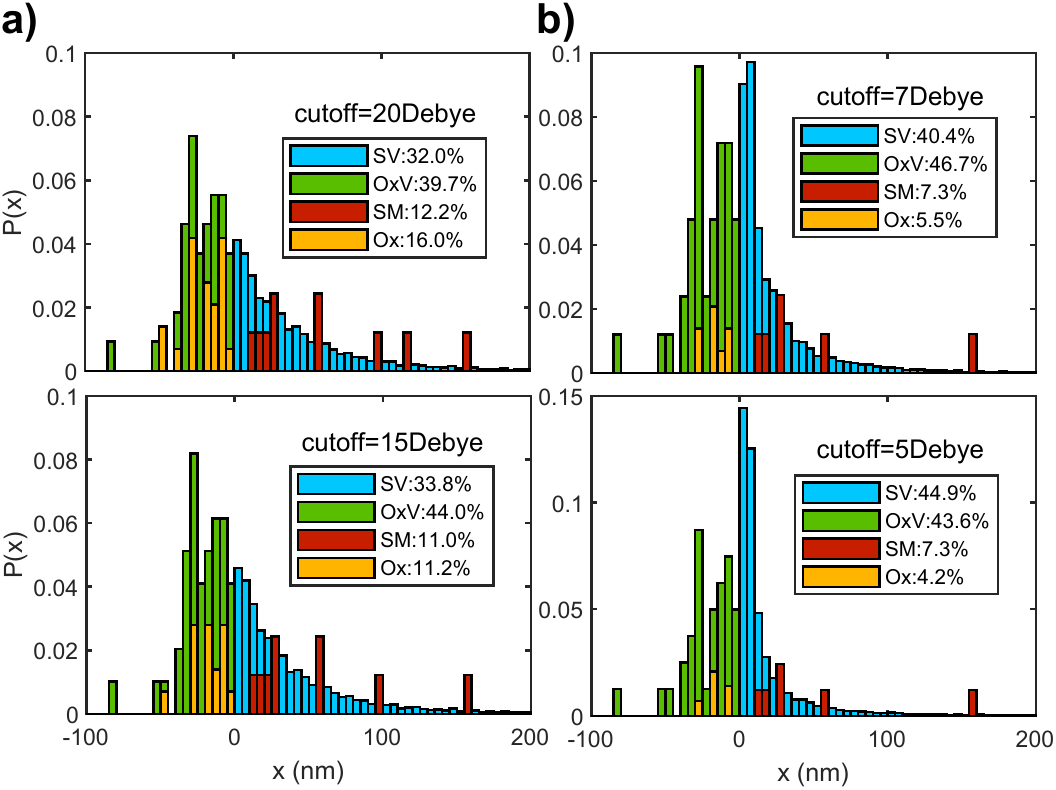}
	\caption{TLS location histograms and interface contributions (legends) deduced for cutoffs \textbf{a} beyond and \textbf{b} below the $\mathrm{10\,D}$ cutoff used in this work and by which the TLS locations are truncated.}
	\label{fig_xHisto_cutoffs_Appendix}
\end{figure}

\noindent\textbf{Cutoff value}\\To discuss the $\mathrm{10\,D}$ cutoff value used in this work to truncate TLS location solutions, we have repeated the analysis for varying cutoffs. Figure \ref{fig_xHisto_cutoffs_Appendix} \textbf{a} shows that a larger cut-off value leads to a larger participation of the Ox interface, which is expected since this interface normally (when the $\mathrm{10\,D}$ cutoff is applied) contains most truncated location solutions. For cutoffs below $\mathrm{10\,D}$ (see \textbf{b}), the histograms become narrower around $x\approx0$ (substrate-metal-vacuum edge), which is expected due to strongest qubit fields in this region. Moreover, smaller cutoffs lead to increased participation of the SV-interface. The reason stems from the method to calculate the probability for a single TLS to reside at a given interface, explained in the main text: TLS locations at film interfaces (OxV, SM, Ox) are degenerate in the TLS orientation and thus they are usually stronger weighted than the orientation-specific solutions at SV-interfaces. On the other hand, in contrast to film interfaces, solutions at SV-interfaces exist for every analyzed TLS even for very small cutoffs. Hence, if all film interface solutions of a TLS happen to be discarded due to a small cutoff value, the SV solutions are maximally weighted which leads to the peak at $x\approx0$ in the SV histograms shown in this figure. To conclude, the $\mathrm{10\,D}$ cutoff is physically motivated: We consider $\mathrm{10\,D}$ as a largest reasonable dipole moment size in solids, and there is no reason to use larger cutoffs. However, smaller cutoffs lead to loss of reasonable solutions and to an unrealistic discontinuity of the location histogram at $x\approx0$.

\subsection{Additional data}
\begin{figure*}
	\includegraphics[width=0.75\linewidth]{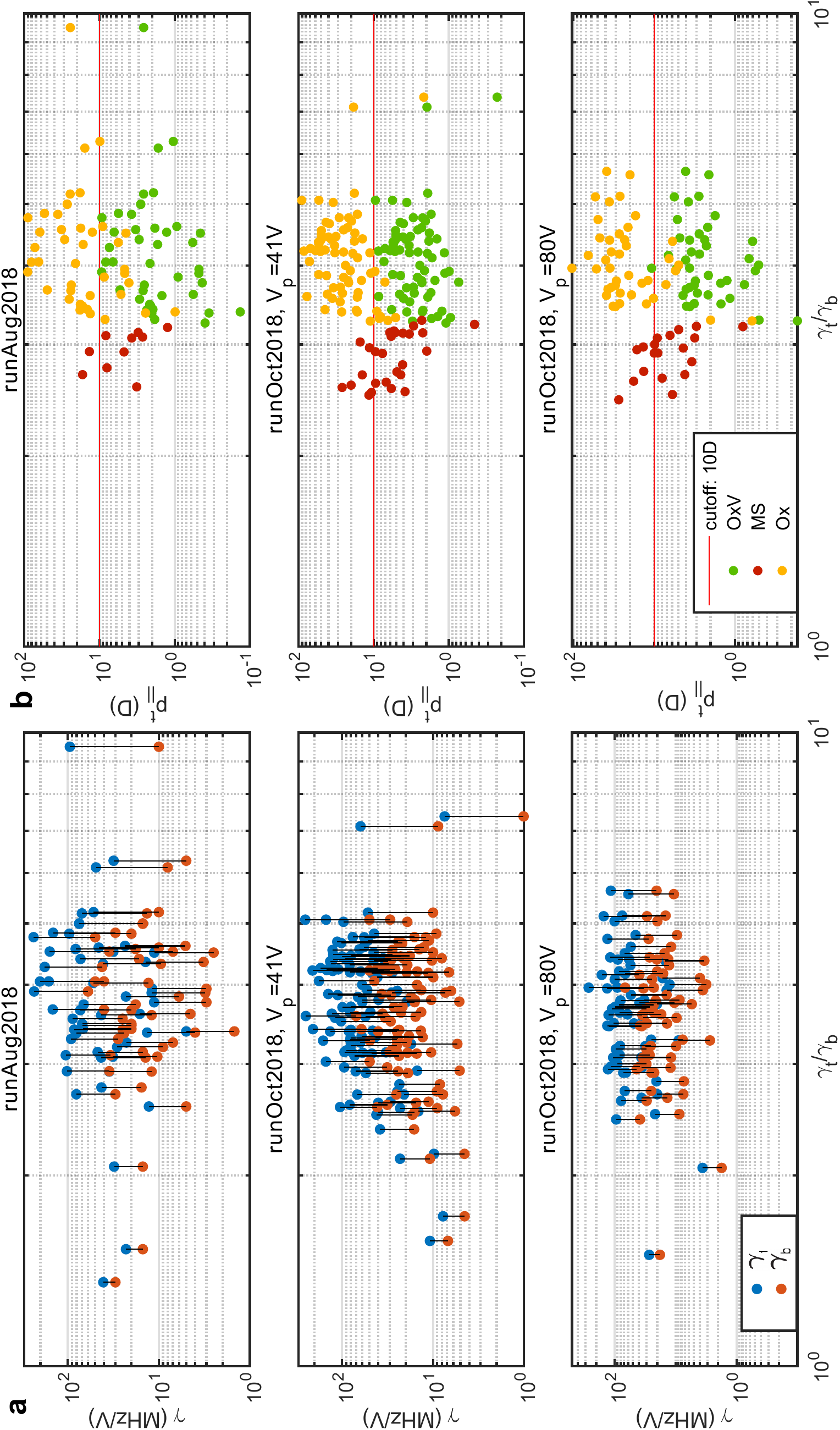}
	\caption{\textbf{Defect tunabilities and dipole moments} \textbf{a} Measured defect tunabilities by top and bottom electrodes $\gamma_\text{t}$ and $\gamma_\text{b}$, respectively, plotted as a function of the tunability ratio $\gamma_\text{t}/\gamma_\text{b}$. From the cryogenic run from October 2018, two data sets exist that were recorded at very different elastic strain applied to the sample chip by a piezo actuator (control voltage $V_\text{p}$), which results in two different set of defects investigated. \textbf{b} Dipole moment components deduced from each defect location and the local electric field. In total, locations of 218 defects were analyzed.}
	\label{fig:gamma_dpar_vs_gamrat}
\end{figure*}

Figure~\ref{fig:gamma_dpar_vs_gamrat}\textbf{a} shows three data sets of defect tunabilities by top and bottom electrodes $\gamma_\text{t}$ and $\gamma_\text{b}$, respectively, plotted against the defect's tunability ratios $\gamma_\text{t}/\gamma_\text{b}$, recorded with Xmon 2 in two cryogenic runs. Fig.~\ref{fig:gamma_dpar_vs_gamrat}\textbf{b} contains dipole moment components deduced from each defect location (cf. Fig.~3) at the film-interfaces (legend). All unrealistic solutions implying a dipole moment larger than 10 Debye are discarded (red line). The data points at tunability ratios below $\sim 2.5$ are missing here since the electric fields at the SM interface were not simulated beyond the distance of $220\,\mathrm{nm}$ from the substrate-metal-vacuum edge. An extrapolation however shows that the electric field at larger distances is so weak that the extracted dipole moments are far above the cutoff value.\\

\begin{figure*}
	\includegraphics[width=\textwidth]{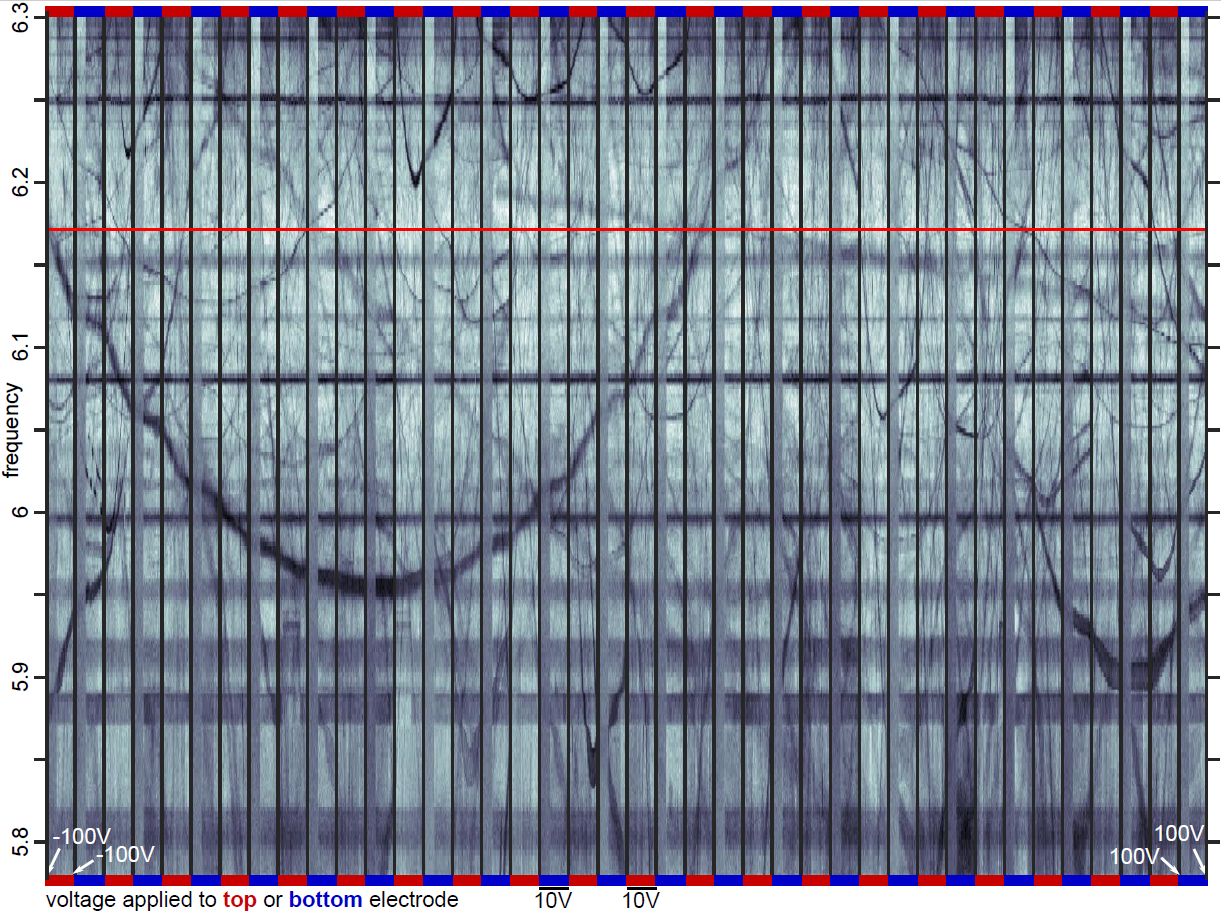}
	\caption{\textbf{Resonances of defects} (dark traces) recorded using swap spectroscopy while sweeping either the voltage applied to the top electrode (red margins) or bottom electrode (blue margins). Each segment spans a range of 10 V, the total range was -100 to +100 V at a step size of 0.14V, the frequency resolution was 3 MHz, and the total measurement duration was 26 hours. The red line indicates the qubit resonance frequency at zero applied magnetic flux. The qubit $T_1$ time was $8.3\,\mathrm{\mu s}$ and the swap duration was $6\,\mathrm{\mu s}$.}
	\label{fig:DC-DC-fastspec-QB2_runAug2018}
\end{figure*}
\begin{figure*}
	\includegraphics[width=\textwidth]{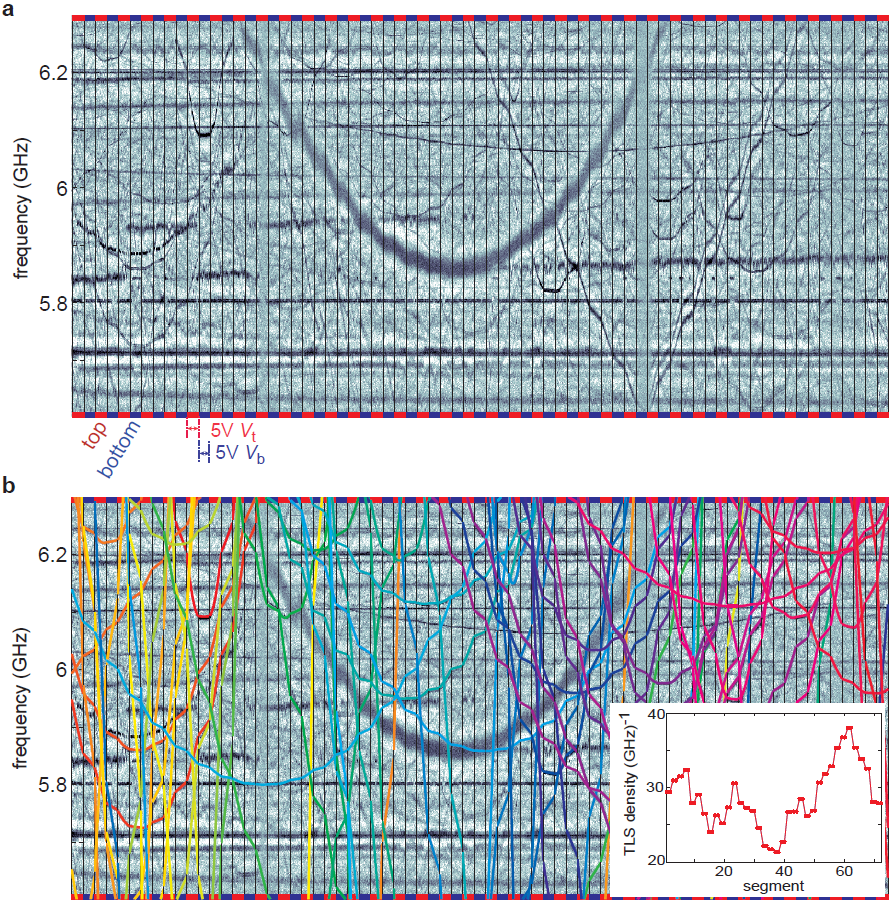}
	\caption{\textbf{Resonances of defects (dark traces)}. \textbf{a} Alternating sweeps of the voltage applied to the top electrode (red margins) and bottom electrode (blue margins). Each segment spans a range of 5 V, the total range was -80V to +100V at a step size of 0.14V, the frequency resolution was 1.5 MHz, and the total measurement duration was 59 hours. \textbf{b} Same data with superimposed fits to segmented hyperbolas (coloured lines). The inset shows the number of detected and fitted defect resonances per GHz in each segment, on average 29 defects/GHz, in total 99 defects.
	\del{XMon2\_Oct18\_44V\_Both.pdf}}
	\label{fig:adddat1}
\end{figure*}

\begin{figure*}
	\includegraphics[width=\textwidth]{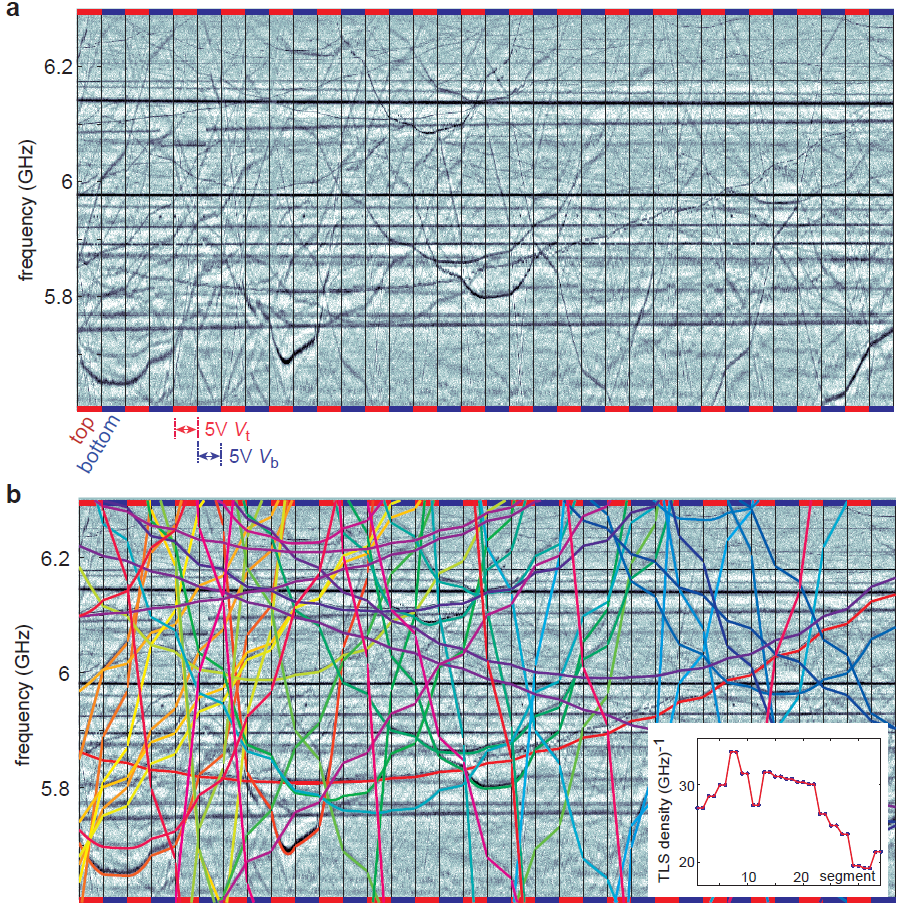}
	\caption{\textbf{Resonances of defects (dark traces)}. Same cool-down as in Fig.~\ref{fig:adddat1}, but taken at a significantly different mechanical strain so that different defects are observed.
 	\textbf{a} Alternating sweeps of the voltage applied to the top electrode (red margins) and bottom electrode (blue margins). Each segment spans a range of 5 V, the total range was 100 to 15 V at a step size of 0.14V, frequency resolution was 1.5 MHz, the total measurement duration was 23 hours. \textbf{b} Same data with superimposed fits to segmented hyperbolas (coloured lines). The inset shows the number of detected and fitted defect resonances per GHz in each segment, on average 27 defects/GHz, in total 63 defects.
	\del{XMon2\_Oct18\_80V\_Both.pdf}
	}
	\label{fig:adddat2}
\end{figure*}

\subsection{Further details}
The here discussed work was part of the PhD studies of Alexander Bilmes at Karlsruhe Institute of Technology (KIT). Further details on the experimental setup, electric field simulations, and data acquisition can be found in his thesis~\cite{Bilmes19}.

\clearpage
\bibliography{TLS_Localization_Biblio}

\end{document}